\documentclass[journal]{IEEEtran}
\usepackage{graphicx}
\usepackage{epstopdf}
\usepackage{cite}
\usepackage{algorithm}
\usepackage{algorithmic}
\usepackage{amsmath}
\usepackage{bm,comment,color,amssymb}
\usepackage{stfloats}
\usepackage{caption}
\usepackage{subcaption}
\newtheorem{remark}{Remark}
\usepackage[colorlinks]{hyperref}

\begin{document}

\title{FAS-assisted NOMA Short-Packet\\Communication Systems}

\author{Jianchao Zheng, Tuo Wu, Xiazhi Lai, Cunhua Pan, Maged Elkashlan, Kai-Kit Wong, \emph{Fellow, IEEE}

\IEEEcompsocitemizethanks{\IEEEcompsocthanksitem \emph{(Corresponding authors: Tuo Wu and Cunhua Pan)}.
\IEEEcompsocthanksitem J. Zheng is with the School of Computer Science and Engineering, Huizhou University, Huizhou 516000, China (E-mail: zhengjch@hzu.edu.cn.). T. Wu and M. Elkashlan are with the School of Electronic Engineering and Computer Science at Queen
Mary University of London, London E1 4NS, U.K. (Email:\{tuo.wu, maged.elkashlan\}@qmul.ac.uk). X. Lai is with the School of Computer Science, Guangdong University of Education, Guangzhou, Guangdong, China (E-mail: xzlai@outlook.com). C. Pan is with the National Mobile Communications Research Laboratory, Southeast University, Nanjing 210096, China. (e-mail: cpan@seu.edu.cn). K. K. Wong is with the Department of Electronic and Electrical Engineering, University College London, WC1E 6BT London, U.K., and also with the Yonsei Frontier Laboratory and the School of Integrated Technology, Yonsei University, Seoul 03722, South Korea (e-mail: kat-kit.wong@ucl.ac.uk).
}

}

\markboth{}
{Zheng \MakeLowercase{\textit{et al.}}:  FAS-assisted NOMA Short-Packet Communication Systems}

\maketitle

\begin{abstract}
In this paper, we investigate a fluid antenna system (FAS)-assisted downlink non-orthogonal multiple access (NOMA) for short-packet communications. The base station (BS) adopts a single fixed antenna, while both the central user (CU) and the cell-edge user (CEU) are equipped with a FAS. Each FAS comprises $N$ flexible positions (also known as ports), linked to $N$ arbitrarily correlated Rayleigh fading channels. We derive expressions for the average block error rate (BLER) of the FAS-assisted NOMA system and provide asymptotic BLER expressions. We determine that the diversity order for CU and CEU is $N$, indicating that the system performance can be considerably improved by increasing $N$. Simulation results validate the great performance of FAS.
\end{abstract}

\IEEEpeerreviewmaketitle

\section{Introduction}
The rapid advances in wireless technologies signals a new chapter in wireless communication. One technology leading to this change is non-orthogonal multiple access (NOMA), which offers improvements in energy and spectral efficiency \cite{DingZ14,IslamS16,DingZ16}.  Recognizing the pressing need for reduced communication delay, Polyanskiy \emph{et al.} introduced  the concept of short-packet communications \cite{Polyanskiy10}, where the block error rate (BLER) has become an important performance metric \cite{Polyanskiy10,Makki14,YuY18}.

Besides, there have been some exciting developments in the field of antenna and radio frequency (RF) technologies. One innovative idea is to use flexible materials such as liquid metal to create unique antenna designs. More recently, pixel-based switchable antennas further provide delay-free reconfiguration of antennas. These novel designs are embraced in the concept of fluid antenna system (FAS) which includes all forms of movable and non-movable flexible-position antennas \cite{Wong-frcmn22}.

In \cite{WongK20}, the ergodic capacity for a single-antenna FAS was first studied while \cite{WongK21} deepened the outage probability analysis. Subsequent work in \cite{WongK22} then offered a closed-form expression that adeptly characterizes the spatial correlation across all the positions (known as ports), directly linking it to the size of the fluid antenna. In \cite{Khammassi-2022}, Khammassi {\em et al.}~proposed better models to account for the impact of spatial correlation. Then \cite{NewW23} revisited the performance analysis of FAS and presented new results to quantify the diversity order of FAS. Broadening the scope of FAS applications, Wong \emph{et al.}~also delved into its potential for multiple access, as elucidated in \cite{WongK222,WongK23}.

Utilizing FAS at mobile devices even with limited space offers a great spatial diversity gain, which matches perfectly with the principles of short-packet communication.  Furthermore, when compared to the traditional orthogonal multiple access (OMA) system, combining FAS with NOMA can further elevate the performance. However, the integration of FAS into NOMA short-packet systems remains unexplored. To fill this research gap, this paper delves into exploring the potential of the FAS-assisted downlink NOMA short-packet system. The main contributions can be summarized as follows:
\begin{itemize}
\item We formulate a FAS-aided downlink NOMA short-packet system model consisting of a base station (BS), a central user (CU), and a cell-edge user (CEU). Both CU and CEU are equipped with fluid antennas, enabling dynamic repositioning amongst $N$ preset locations (or ports). The BS transmits short packets of $N_c$ bits to the CU and $N_e$ bits to the CEU, both with a blocklength of $L$.
\item We theoretically analyze the average BLER at the CU and the CEU based on the Jake's model in \cite{NewW23} representing rich scattering scenarios using both the linear approximation and Gauss-Chebyshev quadrature. Additionally, we present the asymptotic expressions for the BLER by using the first-order Riemann integral approximation.
\item The results indicate that the diversity order for both CU and CEU is $N$. Consequently, a large $N$ can significantly improve the system performance.
\item The simulation results  corroborate the correctness of the derivations of our theoretical analysis.
 \end{itemize}

\section{System Model}
Consider a FAS-assisted downlink NOMA communication system that includes a BS, a CU, and a CEU. All terminals have a single antenna. Both CU and CEU use a fluid antenna to receive signals from the BS while the BS transmits using a fixed antenna. In this system, the BS transmits packets of $N_c$ bits with a blocklength of $L$ to the CU and packets of $N_e$ bits with the same blocklength to the CEU. The fluid antennas at the CU and CEU are assumed to always switch their positions to the optimal position amongst $N$ preset positions within a linear space of size $W\lambda$, where $\lambda$ represents the radiation wavelength. The time delay for port switching is negligible, when pixel-based switchable antennas are considered \cite{Wong-frcmn22}.

Due to the close proximity of the ports within the FAS, they exhibit a significant spatial correlation. Relying on the Jake's model \cite{NewW23}, the spatial correlation between the $m$-th and $n$-th ports can be expressed as
\begin{align}\label{q1}
J_{m,n}=\sigma^2 J_0\left(\frac{2\pi(m-n)W}{N-1}\right),
\end{align}
where $\sigma^2$ represents the large-scale fading effect and $J_0(\cdot)$ denotes the zero-order Bessel function of the first kind.

To better investigate the performance of the overall system, we define the correlation matrix as $\mathbf{J}$. Explicitly, $\mathbf{J}$ can be formulated as
\begin{equation}\label{q2}
   \mathbf{J}=
	\begin{bmatrix}
	 J_{1,1} & \dots & J_{1,N} \\
	 \vdots &  \ddots & \vdots \\
	 J_{N,1} & \dots & J_{N,N}
	\end{bmatrix}.
\end{equation}
For the elements within $\mathbf{J}$ in \eqref{q2}, we assume that $J_{m,n} = J_{n,m}$. Consequently, leveraging the eigenvalue decomposition, the matrix $\mathbf{J}$ can be decomposed as $\mathbf{J} = \mathbf{U \Lambda U}^H$, where $\mathbf{U}$ is an $N \times N$ matrix with its $n$-th column, represented as $\mathbf{u}_n$. Concurrently, $\mathbf{\Lambda} = \text{diag}(\lambda_1, \dots, \lambda_N)$ is an $N \times N$ diagonal matrix, where the $n$-th diagonal entry corresponds to the eigenvalue of $\mathbf{u}_n$. For the sake of analytical convenience, it is assumed that the eigenvalues in $\mathbf{\Lambda}$ are sequenced in a descending order, i.e.,  $\lambda_1 \geq \lambda_2 \geq \cdots \geq \lambda_N$.

Concerning the spatial correlation mentioned above, the complex channel at the $n$-th port for the CU and CEU, denoted as $g^{(c)}_{n}$ and $g^{(e)}_{n}$, can be formulated  as
\begin{align}\label{q5}
   \setlength{\jot}{5pt}
   g^{(c)}_{n}=&\sum_{m=1}^N u_{n,m}\sqrt{\lambda_m}\omega_{cm},\\
   \label{q6}g^{(e)}_{n}=&\sum_{m=1}^N u_{n,m}\sqrt{\lambda_m}\omega_{em},
\end{align}
 where $u_{n,m}$ represents the $(n,m)$-th component of $\mathbf{U}$. Besides, it is assumed that $\omega^{(c)}_{m}=a^{(c)}_{m}+jb^{(c)}_{m}$ and $\omega^{(e)}_{m}=a^{(e)}_{m}+jb^{(e)}_{m}$. The terms $a^{(c)}_{m}$, $b^{(c)}_{m}$, $a^{(e)}_{m}$, and $b^{(e)}_{m}$, ${\forall}m$ are  assumed to be independent and identically distributed (i.i.d.) Gaussian random variables, each with a mean of zero and a variance of $\frac{1}{2}$, respectively.

Furthermore, we assume that the employed FAS within both the CU and the CEU is equipped with a single RF chain. Consequently, at any given time, only one port of the FAS can be active for communication with one user. Hence, the signals received at the $n$-th port for the CU and the CEU can be expressed as
\begin{align}
\label{q3}
y^{(c)}_{n} =& g^{(c)}_{n}d_{c}^{-\frac{a}{2}}\left(\sqrt{\alpha_c P}s_c+\sqrt{\alpha_e P}s_e\right)+z^{(c)}_{n},\\
\label{q4}
y^{(e)}_{n} =& g^{(e)}_{n}d_{e}^{-\frac{a}{2}}\left(\sqrt{\alpha_c P}s_c+\sqrt{\alpha_e P}s_e\right)+z^{(e)}_{n},
\end{align}
where $n \in \{1,2,\dots, N\}$. The terms $z^{(c)}_{n}$ and $z^{(e)}_{n}$ denote the zero-mean complex Gaussian noise at the $n$-th port with a variance of $\delta^2$. The symbols $s_c$ and $s_e$ represent the information symbol transmitted from the BS to the CU and CEU, respectively. We assume that $\mathbb{E}[|s_c|^2]=1$ and $\mathbb{E}[|s_e|^2]=1$. The factors $\alpha_c$ and $\alpha_e$ represent power allocation and satisfy  $\alpha_c + \alpha_e = 1$. Also, $P$ denotes the BS transmit power, while $d_{c}$ and $d_{e}$ represent the distances from the BS to the CU and CEU, respectively, and $a$ is the path loss exponent.

To optimize the system performance, it is assumed that users can dynamically adjust the fluid antenna to be connected to the most advantageous port. For the clarity of analysis, the overall communication performance is assessed based on the maximum values of $g^{(c)}_{n}$ and $g^{(e)}_{n}$, expressed as
\begin{align}
\left|g^{(c)}_{\textrm{FAS}}\right|=&\max\left\{|g^{(c)}_{1}|,|g^{(c)}_{2}|,\dots,|g^{(c)}_{N}|\right\},\label{q7}\\
\left|g^{(e)}_{\textrm{FAS}}\right|=&\max\left\{|g^{(e)}_{1}|,|g^{(e)}_{2}|,\dots,|g^{(e)}_{N}|\right\}.\label{q8}
\end{align}

Upon receiving the signal $y^{(c)}_{n}$, the CU prioritizes decoding the signal $s_e$ by regarding $s_c$ as interference. We can obtain the signal-to-interference-plus-noise ratio (SINR) for decoding $s_e$ at the CU as
\begin{align}\label{q9}
     \gamma_{ce} = \frac{\alpha_e \rho d_c^{-a} |g^{(c)}_{\textrm{FAS}}|^2}{\alpha_c \rho d_c^{-a} |g^{(c)}_{\textrm{FAS}}|^2 + 1},
\end{align}
with $\rho = \frac{P}{\delta^2}$ representing the ratio between the transmit power and noise variance. As given in \cite{Polyanskiy10}, the BLER  for decoding $s_e$ at the CU can be approximated as
\begin{align}\label{q10}
   \epsilon_{ce} \approx  \Psi(\gamma_{ce}, N_e, L),
   \triangleq  Q\left(\frac{C(\gamma_{ce}) - N_e/L}{\sqrt{V(\gamma_{ce})/L}}\right),
\end{align}
where $Q(x) = \frac{1}{\sqrt{2\pi}} \int_{x}^{\infty} e^{-\frac{t^2}{2}} d$, $C(\gamma) = \log_2(1+\gamma)$, and $V(\gamma) = (\log_2 e)^2 \cdot (1-(1+\gamma)^{-2})$ denote the Gaussian Q-function, Shannon capacity formula, and the channel dispersion, respectively.

Upon successfully decoding $s_e$ at the CU, the CU subsequently decodes $s_c$. The instantaneous signal-to-noise ratio (SNR) for decoding $s_c$ can be expressed as
\begin{align}\label{q14}
     \gamma_{cc} = \alpha_c \rho d_c^{-a} |g^{(c)}_{\textrm{FAS}}|^2.
\end{align}
Consequently, the instantaneous BLER when decoding $s_c$ at the CU can be approximated by
\begin{align}\label{q15}
   \epsilon_{cc} \approx  \Psi(\gamma_{cc}, N_c, L),
   \triangleq  Q\left(\frac{C(\gamma_{cc}) - N_c/L}{\sqrt{V(\gamma_{cc})/L}}\right).
\end{align}
Therefore, the BLER of decoding $s_c$ at the CU is given by
\begin{align}\label{q16}
\epsilon_c&=\epsilon_{ce}+(1-\epsilon_{ce})\epsilon_{cc}.
\end{align}

For the CEU, after receiving the signal $y^{(e)}_{n}$, it decodes $s_e$ by treating $s_c$ as interference. The instantaneous SINR for decoding $s_e$ at the CEU can be represented as
\begin{align}\label{q17}
   \gamma_{e}=\frac{\alpha_e\rho d_e^{-a}|g^{(e)}_{\textrm{FAS}}|^2}{\alpha_c\rho d_e^{-a}|g^{(e)}_{\textrm{FAS}}|^2+1}.
\end{align}
According to \cite{Polyanskiy10}, the BLER for decoding $y_e$ at the CEU is found as
\begin{align}\label{q18}
   \epsilon_{e} \approx \Psi(\gamma_{e},N_e,L),
   \triangleq  Q\left(\frac{ C(\gamma_{e})-N_e/L}{\sqrt{V(\gamma_{e})/L}}\right).
\end{align}

\section{Theoretical Analysis of Average BLER}
In this section, we will derive the theoretical expression of the average BLER to decode $s_e$ at the CEU and that to decode $s_c$ at the CU. The  theoretical analysis of the average BLER to decode $s_c$ at the CU is obtained by
\begin{align}\label{q19}
  \mathbb{E} [\epsilon_c]=&\mathbb{E}[\epsilon_{cc}]+\mathbb{E}[\epsilon_{ce}]-\mathbb{E}[\epsilon_{cc}\epsilon_{ce}],
\end{align}
where $\mathbb{E}[\epsilon_{cc}]$ is expressed as
\begin{align}\label{q20}
\mathbb{E}\left[\epsilon_{cc}\right]\approx\int_{0}^{\infty}\Psi(\gamma_{cc},N_c,L)f_{\gamma_{cc}}(t)dt.
\end{align}
To further derive $\mathbb{E}\left[\epsilon_{cc}\right]$, we first introduce the linear approximation  of $\Psi(\gamma_{cc},N_c,L)$ as follows \cite{YuY18,Makki14}:
\begin{align}\label{q21}
&\Psi(\gamma_{cc},N_c,L)\nonumber\\&=\left\{
\begin{array}{ll}
1, & \gamma_{cc}\leq v_{N_c,L},\\
\frac{1}{2}-\delta_{N_c,L}\sqrt{L}(\gamma_{cc}-\beta_{N_c,L}), &v_{N_c,L}<\gamma_{cc}<u_{N_c,L},\\
0, & \gamma_{cc}\geq u_{N_c,L},
\end{array}
\right.
\end{align}
where $\beta_{N_c,L}=2^\frac{N_c}{L}-1$, $\delta_{N_c,L}=(2\pi(2^\frac{2N_c}{L}-1))^{-\frac{1}{2}}$, $v_{N_c,L}=\beta_{N_c,L}-\frac{1}{2}\delta_{N_c,L}^{-1}L^{-\frac{1}{2}}$, and $u_{N_c,L}=\beta_{N_c,L}+\frac{1}{2}\delta_{N_c,L}^{-1}L^{-\frac{1}{2}}$.
Substituting \eqref{q21} into \eqref{q20}, we have
\begin{equation}\label{q26}
\mathbb{E}\left[\epsilon_{cc}\right]\approx\delta_{N_c,L}\sqrt{L}\int_{v_{N_c,L}}^{u_{N_c,L}}F_{\gamma_{cc}}(\tau)d\tau,
\end{equation}
where $F_{\gamma_{cc}}(\tau)$ is the cumulative density function (CDF) of $\gamma_{cc}$. However, direct calculation of the integral for $F_{\gamma_{cc}}(\tau)$ proves to be challenging. To further derive the expression of $\mathbb{E}\left[\epsilon_{cc}\right]$, we introduce the following lemma.

\begin{figure*}[!t]
    \centering
    \begin{subfigure}[b]{0.49\linewidth}
        \includegraphics[width=\linewidth]{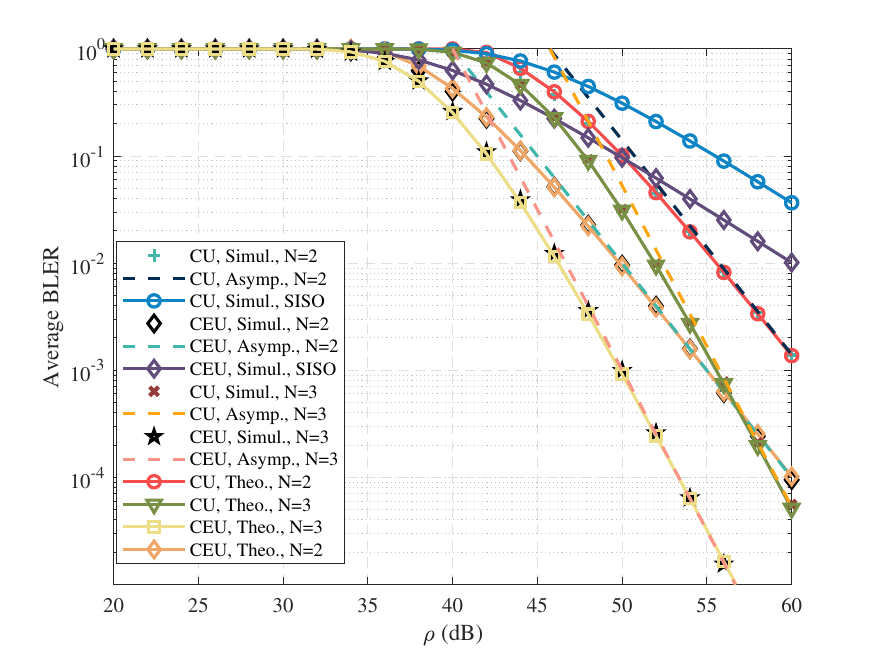}
        \caption{Average BLER versus $\rho$}\label{fig:subfig_a}
    \end{subfigure}
    \hfill
    \begin{subfigure}[b]{0.49\linewidth}
        \includegraphics[width=\linewidth]{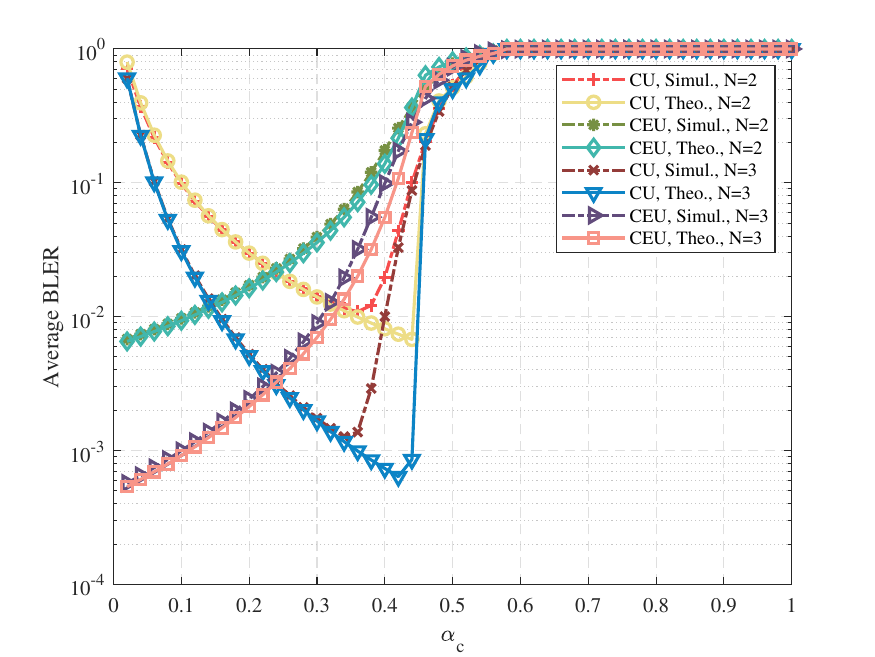}
        \caption{Average BLER versus $\alpha_c$}\label{fig:subfig_b}
    \end{subfigure}
\caption{\small  (a) Parameters: $L = 100$, $N_c = 300$ bits, $N_e = 100$ bits, $\{N=2, W=5\}$ or $\{N=3, W=10\}$, $\alpha_c = 0.1$, $\alpha_e = 0.9$. (b) Parameters: $L = 100$, $N_c = 300$ bits, $N_e = 100$ bits, $\rho=50$ dB, $\{N=2, W=5\}$ or $\{N=3, W=10\}$.}\label{fig:combined_fig}
\end{figure*}

\emph{Lemma 1:} Let $D = \det(\mathbf{J})$, $ P_t = \frac{(-K_{m,n})^{s_t^*}}{{s_t^*}!D}$, and $P_n = \frac{1}{2}\left(\frac{K_{n,n}}{D}\right)^{-\frac{\bar{s}_n}{2}-\frac{1}{2}}\left[\Gamma\left(\frac{1+\bar{s}_n}{2}\right)-\Gamma\left(\frac{1+\bar{s}_n}{2},\frac{K_{n,n}d_c^{a}\tau}{\alpha_c\rho D}\right)\right]$, the CDF of ${\gamma_{cc}}$ is given by
\begin{align}\label{q27}
   F_{\gamma_{cc}}(\tau) = \sum_{s_1=0}^{s_0}... \sum_{s_T=0}^{s_T}\frac{g(\mathbf{s}^*)}{\pi^N D} \prod_{t=1}^T P_t \prod_{n=1}^N P_n,
   \end{align}
where
\begin{align}\label{q28}
   g(\mathbf{s}^*)=\frac{1}{2}^{\sum_{t=1}^T s_t^*}\sum_{\mathbf{v}\in \mathcal{V}}\left[\prod_{t=1}^T\binom {s_t^*} {v_t}\right]\left(2\pi\right)^N\prod_{i=1}^N \mathbf{1}_{\{\Delta_i=0\}},
\end{align}
and $\bar{s}_n=\sum_{i=1}^N S_{n,i}^*+\sum_{i=1}^{n-1} S_{i,n}^*+1$, $S_{i,n}^*$ denotes the $(i,n)$-th entry of $\mathbf{S}^*$. Besides, $\mathbf{S}^*$ is given by
\begin{equation}\label{q29}
   \mathbf{S}^*=
	\begin{bmatrix}
	 0      & s_1^*  & \dots & s_{N-1}^* \\
	        & \vdots & \ddots & \vdots\\
           & 0     & \dots & s_T^*
	 \end{bmatrix},
\end{equation}
where $s_t^* = s_t - s_{t+1}$ with $s_{T+1}=0$. Additionally, $s_0$ is a finite constant which must be sufficiently large for the approximation to be accurate. The index $t$ is defined as $t=n+(m-1)N-\frac{m(m+1)}{2}$, where $m<n$. The indices $m$ and $n$ can be derived from $t$, such that $m$ is the minimal integer $m'$ satisfying the condition $\sum_{i=1}^{m'}(N-i) > t$, and $n=t-(m-1)N+\frac{m(m+1)}{2}$.

Furthermore, $\mathbf{v}=[v_1,\dots,v_T]^T$ and $\mathcal{V}$ represents the set of all possible permutations. The variable $\Delta_i$ is given by $\Delta_i=\sum_{n=1}^N G_{i,n}-\sum_{n=1}^N G_{n,i}-G_{i,i}$. The entry $K_{m,n}$ is the $(m,n)$-th element of $\mathbf{K}$, where $\mathbf{K}$ is the co-factor of $\mathbf{J}$. Similarly, $G_{m,n}$ is the $(m,n)$-th element of $\mathbf{G}$, which is defined as
\begin{equation}\label{q30}
   \mathbf{G}=
	\begin{bmatrix}
	 0      & \gamma_1  & \dots & \gamma_{N-1} \\
	        & \vdots & \ddots & \vdots\\
           & 0     & \dots & \gamma_T
	 \end{bmatrix},
\end{equation}
where $\gamma_t = 2v_t - s_t^*$ and $\gamma_t \in  \mathbb{Z}$. Here, $\Gamma(\cdot)$ stands for the Gamma function, while $\Gamma(a,b)$ represents the upper incomplete gamma function.

\begin{proof}
See Appendix A.
\end{proof}

Following \emph{Lemma 1}, we can substitute \eqref{q27} into \eqref{q26} and utilize the Gauss-Chebyshev quadrature \cite{Hildebrand87}, which yields  the following result
\begin{align}\label{q31}
  \mathbb{E}\left[\epsilon_{cc}\right]\approx \frac{\pi}{2U_p}\sum_{p=1}^{U_p}\sqrt{1-\eta_p^2} R,
\end{align}
where $ R = \sum_{s_1=0}^{s_0}\cdots \sum_{s_T=0}^{s_T}\frac{g(\mathbf{s}^*)}{\pi^N D} \prod_{t=1}^T P_t \prod_{n=1}^N P_n
$, and $U_p$ represents the complexity-accuracy tradeoff parameters, $ \eta_p=\cos\left(\frac{(2p-1)}{2U_p}\pi\right)$, and $y_{pc}=\frac{\eta_p}{2\delta_{N_c,L}\sqrt{L}}+\beta_{N_c,L}$.

Drawing a parallel, by invoking \emph{Lemma 1}, when $\tau\leq\frac{\alpha_e}{\alpha_c}$,  we have the CDF of $\gamma_{ce}$ given by
\begin{align}\label{q33}
   F_{\gamma_{ce}}(\tau) = \sum_{s_1=0}^{s_0}\cdots \sum_{s_T=0}^{s_T}\frac{g(\mathbf{s}^*)}{\pi^N D} \prod_{t=1}^T P_t \prod_{n=1}^N P'_n,
\end{align}
where
\begin{multline}
P'_n = \frac{1}{2}\left(\frac{K{n,n}}{D}\right)^{-\frac{\bar{s}_n}{2}-\frac{1}{2}}\times\\
\left[\Gamma\left(\frac{1+\bar{s}_n}{2}\right)-\Gamma\left(\frac{1+\bar{s}n}{2},\frac{K{n,n}d_c^{a}\tau}{(\alpha_e\rho-\alpha_c\rho\tau) D}\right)\right].
\end{multline}

Similarly with $\mathbb{E}\left[\epsilon_{cc}\right]$, we have
\begin{equation}\label{q34}
   \mathbb{E}[\epsilon_{ce}] \approx \delta\sqrt{m}\int_{v}^{u} F_{\gamma_{ce}}(\tau)d\tau,
\end{equation}
where $\delta =(2\pi(2^\frac{2N_e}{L}-1))^{-\frac{1}{2}}$, $v = \beta-\frac{1}{2}\delta^{-1}L^{-\frac{1}{2}}$, and $\beta = 2^\frac{N_e}{L}-1$, respectively.
Consequently, by substituting \eqref{q33} into \eqref{q34} and assuming $Q = \frac{g(\mathbf{s}^*)}{\pi^N D}$, we obtain
\begin{align}\label{q36}
   \mathbb{E}\left[\epsilon_{ce}\right] \approx \frac{\pi}{2U_p}\sum_{p=1}^{U_p}\sqrt{1-\eta_p^2} \times \sum_{s_1=0}^{s_0}\cdots\sum_{s_T=0}^{s_T} Q \prod_{t=1}^T P_t \prod_{n=1}^N P'_n,
\end{align}
 where
\begin{align}\label{q37}
   y_{pe}=\frac{\eta_p}{2\delta_{N_e,L}\sqrt{L}}+\beta_{N_e,L},
\end{align}
for $y_{pe}\leq\frac{\alpha_e}{\alpha_c}$; otherwise, $\mathbb{E}\left[\epsilon_{ce}\right]=1$.

Utilizing \eqref{q36} and \eqref{q31} and following \cite{LaiX19}, we have the average BLER at the CU formulated as
\begin{align}\label{q38}
  \mathbb{E} [\epsilon_c]\geq\textrm{max} \left\{ \mathbb{E}[\epsilon_{cc}],\mathbb{E}[\epsilon_{ce}]\right\}.
\end{align}

For the derivation of $\mathbb{E}[\epsilon_{e}]$, when $\tau<\frac{\alpha_e}{\alpha_c}$, the CDF of $\gamma_{e}$ can be expressed as
\begin{align}\label{q39}
   F_{\gamma_{e}}(\tau) = \sum_{s_1=0}^{s_0}\cdots\sum_{s_T=0}^{s_T} S \prod_{t=1}^T R_t \prod_{n=1}^N T_n U_n,
\end{align}
where $R_t = \frac{(-K_{m,n})^{s_t^*}}{{s_t^*}!D}$, $S = \frac{g(\mathbf{s}^*)}{\pi^N D}$, $T_n = \frac{1}{2}\left(\frac{K_{n,n}}{D}\right)^{-\frac{\bar{s}_n}{2}-\frac{1}{2}}$, and $U_n = \Gamma\left(\frac{1+\bar{s}_n}{2}\right) - \Gamma\left(\frac{1+\bar{s}_n}{2},\frac{K_{n,n}d_e^{a}\tau}{(\alpha_e\rho-\alpha_c\rho\tau)D}\right)$. Otherwise, $F_{\gamma_{e}}(\tau)=1$. Similar to $\mathbb{E}\left[\epsilon_{ce}\right]$, we have
\begin{equation}\label{q40}
   \mathbb{E}\left[\epsilon_{e}\right]\approx\delta_{N_e,L}\sqrt{m}\int_{v_{N_e,L}}^{u_{N_e,L}}F_{\gamma_{e}}(\tau)d\tau.
\end{equation}
Substituting \eqref{q36} into \eqref{q37}, we obtain
\begin{align}\label{q41}
  \mathbb{E}\left[\epsilon_{e}\right] \approx \frac{\pi}{2U_p}\sum_{p=1}^{U_p} \sqrt{1-\eta_p^2} \sum_{s_1=0}^{s_0}\cdots\sum_{s_T=0}^{s_T} S \prod_{t=1}^T R_t \prod_{n=1}^N T_n U_n,
\end{align}
for $y_{pe}<\frac{\alpha_e}{\alpha_c}$; otherwise, $\mathbb{E}\left[\epsilon_{e}\right]=1$.

\section{Asymptotic Analysis}
From \cite{NewW23}, the CDF of $|g^{(c)}_{\textrm{FAS}}|$ and $|g^{(e)}_{\textrm{FAS}}|$ at high SNR is given by
\begin{align}\label{q42}
   F_{|g^{(c)}_{\textrm{FAS}}|\mbox{ or }|g^{(e)}_{\textrm{FAS}}|}(x)=\frac{1}{D}x^{2N}+o\left(\frac{1}{\rho^N}\right).
\end{align}
Consequently, by using the first-order Riemann integral approximation
\begin{align}\label{q43}
   \int_a^b f(\tau)d\tau=\frac{b-a}{2}f\left(\frac{b+a}{2}\right),
\end{align}
the asymptotic expression of $\mathbb{E}\left[\epsilon_{cc}\right]$, $\mathbb{E}\left[\epsilon_{ce}\right]$ and $\mathbb{E}\left[\epsilon_{e}\right]$ for high SNR can be approximated as
\begin{align}\label{q44}
   \mathbb{E}\left[\epsilon_{cc}\right]\approx& \frac{1}{D}\left(\frac{d_c^a}{\alpha_c\rho}\beta_{N_c,L}\right)^{N},\\
   \mathbb{E}\left[\epsilon_{ce}\right]\approx& \frac{1}{D}\left(\frac{d_c^a\beta_{N_e,L}}{\alpha_e\rho-\alpha_c\rho\beta_{N_e,L}}\right)^{N},\\
   \mathbb{E}\left[\epsilon_{e}\right]\approx& \frac{1}{D}\left(\frac{d_e^a\beta_{N_e,L}}{\alpha_e\rho-\alpha_c\rho\beta_{N_e,L}}\right)^{N},
\end{align}
for $\beta_{N_e,L}\leq \frac{\alpha_e}{\alpha_c}$; otherwise, $\mathbb{E}\left[\epsilon_{ce}\right]=1$ and $\mathbb{E}\left[\epsilon_{e}\right]=1$.
\begin{remark}
As we can see from \eqref{q44}, we can deduce a noteworthy observation that the diversity order for both the CU and the CEU is directly  defined by $N$. This insight suggests that increasing $N$ can play a significant role in enhancing the overall performance and robustness of the system.
\end{remark}

\section{Numerical Results}
Here, we provide numerical results to validate the accuracy of the analytically derived BLERs. For our simulations, we adopt the following parameters: $d_c = 5 \text{ m}$, $d_e = 10 \text{ m}$, $a = 3.9$, and $U_p = 10$. It is imperative to highlight that as we increase the value of $U_p$, our results show even better accuracy. We set the blocklength to $L = 100$, and the respective number of data bits allocated for CU and CEU are $N_c = 300 \text{ bits}$ and $N_e = 100 \text{ bits}$, following the setup suggested by \cite{LaiX19}.

In Fig.~\ref{fig:combined_fig}(a), the relationship between the average BLER and SNR $\rho$  is depicted for both CU and CEU under various parameters. We have set the power allocation coefficients to $\alpha_c = 0.1$ and $\alpha_e = 0.9$   \cite{LaiX19}, with either $N = 3$ or $N = 4$. ``CU, Simul.'', ``CU, Theo.'', and ``CU, Asymp.'' represent the simulated, theoretical, and asymptotic average BLER outcomes for CU. Similarly, ``CEU, Simul.'', ``CEU, Theo.'', and ``CEU, Asymp.'' denote the  related results for CEU. For comparison, we also depict the simulated average BLER for CU and CEU in a single antenna setup without FAS, labeled as ``CU, SISO, Simul.'' and ``CEU, SISO, Simul.''. From Fig.~\ref{fig:combined_fig}(a), it is seen that as SNR $ \rho $ increases, there is a remarkable decline in the average BLER. Also, the theoretical lines closely match the simulated results. A comparison with the non-FAS highlights the improved performance of the FAS-assisted NOMA short-packet system. 

In Fig.~\ref{fig:combined_fig}(b), the average BLER is shown as a function of $\alpha_c$ for a fixed $\rho = 40$ dB and $N = 3$. The results  in Fig.~\ref{fig:combined_fig}(b) demonstrate the consistency between the theoretical analysis and the simulated outcomes for the average BLERs of both CU and CEU, confirming the robustness of our proposed model.

\section{Conclusion}
This paper derived the average BLER of the FAS-assisted downlink NOMA short-packet communication system.  Numerical results showed that the average BLERs of both the CU and CEU are nearly identical. Moreover, the FAS-assisted NOMA short-packet system outperforms its non-FAS counterpart, demonstrating great performance enhancement.

\section{Appendix A}
The joint CDF of $|g_{c1}|$, $|g_{c2}|$,$\dots$,$|g_{cN}|$ is given by \cite{NewW23}
\begin{align}\label{aq1}
F&_{|g_{c1}|,|g_{c2}|,\dots,|g_{cN}|}(r_1,r_2,\dots,r_N)\nonumber\\
=&\sum_{s_1=0}^{s_0}\sum_{s_2=0}^{s_1}\cdots\sum_{s_T=0}^{s_T}\frac{g(\mathbf{s}^*)}{\pi^N\det(\mathbf{J})}\prod_{t=1}^T\frac{(-K_{m,n})^{s_t^*}}{{s_t^*}!\det(\mathbf{J})}\times\nonumber\\
&\prod_{n=1}^N\frac{1}{2}\left(\frac{K_{n,n}}{\det(\mathbf{J})}\right)^{-\frac{\bar{s}_n}{2}-\frac{1}{2}}\times\nonumber\\
&\left[\Gamma\left(\frac{1+\bar{s}_n}{2}\right)-\Gamma\left(\frac{1+\bar{s}_n}{2},\frac{K_{n,n}r_n^2}{\det(\mathbf{J})}\right)\right].
\end{align}
With the joint cdf of $|g_{c1}|$, $|g_{c2}|$,\dots,$|g_{cN}|$, we use the following expression to compute the CDF of $\gamma_{cc}$ as
\begin{align}\label{aq2}
F_{\gamma_{cc}}(\tau)=&\Pr\left(\gamma_{cc}<\tau\right).
\end{align}
Adding \eqref{q14} into \eqref{aq2}, we have
\begin{align}
   F_{\gamma_{cc}}(\tau)=&\Pr\left(|g^{(c)}_{\textrm{FAS}}|<\sqrt{\frac{ d_c^{a}\tau}{\alpha_c\rho }}\right)\nonumber\\
   =&\Pr\left(|g_{c1}|<\sqrt{\frac{d_c^{a}\tau}{\alpha_c\rho}},\dots,|g_{cN}|<\sqrt{\frac{d_c^{a}\tau}{\alpha_c\rho}}\right).
\end{align}
The CDF expression can be found by substituting $r_1=r_2=\cdots=r_N=\sqrt{\frac{d_c^{a}\tau}{\alpha_c\rho}}$ into the joint CDF \eqref{aq1}, which complete the proof.

\end{document}